\documentclass[reprint,
 superscriptaddress,
nofootinbib,
 amsmath,amssymb,
 aps,
pra,
]{revtex4-2}

\usepackage[utf8]{inputenc} % allow utf-8 input
\usepackage[T1]{fontenc}    % use 8-bit T1 fonts
\usepackage{hyperref}       % hyperlinks
\usepackage{url}            % simple URL typesetting
\usepackage{booktabs}       % professional-quality tables
\usepackage{amsfonts}       % blackboard math symbols
\usepackage{nicefrac}       % compact symbols for 1/2, etc.
\usepackage{microtype}      % microtypography
\usepackage{graphicx}
\usepackage{xcolor, etoolbox}
\usepackage{amsmath,amsfonts,amsthm} % Math packages
\usepackage{mathtools}
\usepackage{ragged2e}
\usepackage{braket}      % Dirac Bra-Ket notation
\usepackage[english]{babel} % English language/hyphenation
\usepackage{enumitem}
\usepackage{xspace}
\usepackage{dsfont}
\usepackage{bm}% bold math
\usepackage{bbm}% bold math
\usepackage{dcolumn}% Align table columns on decimal point
\usepackage{mathrsfs}
\usepackage{todonotes}
\newcommand{\Tr}{\text{Tr}}

\newcommand{\eq}[1]{Eq.~(\ref{#1})}
\newcommand{\fig}[1]{Fig.~\ref{#1}}

\hypersetup{
    colorlinks,
    linkcolor=blue,
    citecolor=blue,
    urlcolor=blue,
    breaklinks=true 
}

\DeclarePairedDelimiter\abs{\lvert}{\rvert}

\theoremstyle{remark}

\begin{document}

\title{Entropic entanglement criteria in phase space}

\author{Stefan Floerchinger}
    \email{stefan.floerchinger@thphys.uni-heidelberg.de}
    \affiliation{Institut f\"{u}r Theoretische Physik, Universit\"{a}t Heidelberg, Philosophenweg 16, 69120 Heidelberg, Germany}
\author{Martin G\"{a}rttner}
    \email{martin.gaerttner@kip.uni-heidelberg.de}
    \affiliation{Institut f\"{u}r Theoretische Physik, Universit\"{a}t Heidelberg, Philosophenweg 16, 69120 Heidelberg, Germany}
    \affiliation{Physikalisches Institut, Universit\"{a}t Heidelberg, Im Neuenheimer Feld 226, 69120 Heidelberg, Germany}
    \affiliation{\mbox{Kirchhoff-Institut f\"{u}r Physik, Universit\"{a}t Heidelberg,
    Im Neuenheimer Feld 227, 69120 Heidelberg, Germany}}
\author{Tobias Haas}
    \email{t.haas@thphys.uni-heidelberg.de}
    \affiliation{Institut f\"{u}r Theoretische Physik, Universit\"{a}t Heidelberg, Philosophenweg 16, 69120 Heidelberg, Germany}
\author{Oliver R. Stockdale}
    \email{oliver.stockdale@kip.uni-heidelberg.de}
    \affiliation{\mbox{Kirchhoff-Institut f\"{u}r Physik, Universit\"{a}t Heidelberg,
    Im Neuenheimer Feld 227, 69120 Heidelberg, Germany}}

\begin{abstract}
We derive entropic inseparability criteria for the phase space representation of quantum states. In contrast to criteria involving differential entropies of marginal phase space distributions, our criteria are based on a joint distribution known as the Husimi $Q$-distribution. This distribution is experimentally accessible in cold atoms, circuit QED architectures and photonic systems and bears practical advantages compared to the detection of marginals. We exemplify the strengths of our entropic approach by considering several classes of non-Gaussian states where second-order criteria fail. We show that our criteria certify entanglement in previously undetectable regions highlighting the strength of using the Husimi $Q$-distribution for entanglement detection.
\end{abstract}

\maketitle

\section{Introduction}
Entanglement is the distinguishing feature of quantum systems and its detection is critical for characterizing them~\cite{Horodecki2009}. Fundamental problems such as the thermalization of isolated quantum systems and characterizing quantum phase transitions strongly rely on a deep understanding of how entanglement manifests and evolves~\cite{Kaufman2016,Osborne2002}. However, a central problem in studying entanglement is being able to derive experimentally accessible witnesses to detect it~\cite{Guehne2009,Plenio2007}. The complexity of such a problem depends upon the Hilbert space size, such that continuous variable systems with an infinite dimensional Hilbert space pose a particular challenge.

For continuous quantum variables, many entanglement criteria rely on measuring the second-order moments of two marginal distributions~\cite{Simon2000,Duan2000,Mancini2002,Giovannetti2003,Guehne2004,Hyllus2006}. These criteria are most powerful when the state is Gaussian but are often insensitive elsewhere~\cite{Weedbrook2012,Serafini2017}. This can pose significant problems as there are many important classes of highly entangled non-Gaussian states that cannot be witnessed by second-order criteria~\cite{Rodo2008,Dong2008,Strobel2014}. 

To capture higher-order moments of measured distributions~\cite{Shchukin2005,Shchukin2005b}, one can use differential entropies of measured marginal distributions~\cite{Bialynicki-Birula1975,Maassen1988}. Differential entropies reach beyond the scope of second-order criteria since they are a functional of the full probability density function. Examples include criteria that rely on entropic uncertainty relations~\cite{Coles2017,Hertz2019,Bialynicki-Birula2011,Walborn2009,Saboia2011} as well as the complexity based criterion~\cite{Huang2013}. Other approaches are predicated on entropic uncertainty relations with (quantum) memory~\cite{Frank2013,Furrer2014} or as entropic steering inequalities~\cite{Walborn2011,Chowdhury2014}. Entropic criteria have also been derived and experimentally tested for discrete variables~\cite{Guehne2004b,Li2011,Schneeloch2019,Berta2010,Bergh2021a,Bergh2021b}.

One disadvantage to these methods is that measuring marginals of a distribution is often costly and impractical as it requires angle tomography. This is particularly difficult in ultracold quantum gas experiments where statistics are limited~\cite{Phelps2020}. In this work, we take a conceptually different approach and characterize the inseparability of a given bipartite state not by its marginal distributions, but by its joint probability distribution. This distribution, known as the Husimi $Q$-distribution, is a quasiprobability distribution that contains the full information about the state~\cite{Husimi1940,Schleich2001,Mandel2013,Lee1995,Cartwright1976}. Unlike the Wigner function, it is non-negative and hence has an associated entropy known as the Wehrl entropy~\cite{Wehrl1978,Wehrl1979}. 

%\red{here give some insight into the $Q$ function and why it has been overlooked}

The detection of marginal distributions (via Wigner) and the simultaneous detection of phase space variables (via Husimi) for quantum state tomography are considered as two complementary, but in principle equally powerful, approaches~\cite{WELSCH1999}. For jointly measured observables, the full information about correlations between different directions in phase space is simultaneously available; for sequentially measured marginals, a direction needs to be preselected. This suggests the Husimi $Q$-distribution offers unexplored opportunities to derive entanglement witnesses for systems where experimental statistics are limited~\footnote{The Wehrl entropy in the context of entanglement has only been discussed to define an entanglement monotone~\cite{Mintert2004}.}. 

Crucially, the Husimi $Q$-distribution (and hence the Wehrl entropy) is an experimentally accessible quantity that can be measured via tomographic methods~\cite{Landon2018}. This is well-established within quantum optics and has been realized in experiments~\cite{Noh1991,Noh1992,Leonhardt1993}. Recently, the measurement of Husimi $Q$-distributions has been demonstrated on a variety of other platforms, including ultracold Bose gases~\cite{Kunkel2019,Kunkel2021}, atoms in optical cavities \cite{haas2014, barontini2015}, and circuit QED architectures~\cite{kirchmair2013}, showing practical advantages with respect to the detection of marginals.

Here, we derive entanglement criteria in terms of entropies of the Husimi $Q$-distribution. We show that these criteria are stronger than previously known ones for certain classes of states within the non-Gaussian regime. We discuss various experimental platforms where we expect our criteria could be most powerful in terms of implementation and state detection.

\textit{Notation}---We set $\hbar = 1$ and disregard operator hats. We use capital letters for quantum operators $X_j$ and small letters for their corresponding eigenvalues $x_j$ and eigenvectors $\ket{x_j}$.

\section{Husimi $\mathbf{\textit{Q}}$-distribution and Wehrl entropy}
We consider a set of continuous quantum variables $X_j$ and $P_j$ that fulfill the bosonic commutation relation $[X_j, P_k] = i \delta_{jk}$ where the subindices denote the two subsystems $j, k \in \{1,2 \}$. A canonical transformation can make rotations in the local phase spaces by angles $\vartheta_j$, 
\begin{equation}
    \begin{pmatrix}
    R_j \\ S_j
    \end{pmatrix} = 
    \begin{pmatrix}
    \cos \vartheta_j & \sin \vartheta_j \\
    - \sin \vartheta_j & \cos \vartheta_j
    \end{pmatrix}
    \begin{pmatrix}
    X_j \\ P_j
    \end{pmatrix}.
    \label{eq:RotatedVariables}
\end{equation}
To a set of (possibly rotated) position and momentum operators $R_j$ and $S_j$, we define annihilation operators $A_j = (R_j + i S_j)/\sqrt{2}$ such that coherent states are their eigenstates $A_j \ket{\alpha} = \alpha_j \ket{\alpha}$. The complex eigenvalues $\alpha_j$ are parameterized as $\alpha_j = (r_j + i s_j)/\sqrt{2}$. One can associate a positive operator-valued measure (POVM) to pure coherent state projectors of the form $E_{\alpha} = \ket{\alpha} \bra{\alpha}$. This POVM is experimentally accessible through tomographic schemes~\cite{Landon2018} or the heterodyne detection protocol commonly used in quantum optical systems~\cite{Weedbrook2012,Schleich2001,Mandel2013}, which corresponds to measuring the state $\rho$ in the pure coherent state basis.

The distribution associated to the measurement outcomes for a set of conjugate quantum variables is the global Husimi $Q$-distribution 
\begin{equation}
    Q(r_1, s_1, r_2, s_2)  = \Tr \{\rho E_{\alpha} \} = \braket{\alpha | \rho | \alpha}.
    \label{eq:HusimiQDefinition}
\end{equation}
Due to the non-orthogonality of the coherent states, the Husimi $Q$-distribution is a quasi-probability distribution in phase space. The readout angles $\vartheta_i$ can then be chosen in a \textit{post-measurement} analysis unlike measuring marginals of the Wigner distribution where these angles need to be preselected. Also, the Husimi $Q$-distribution is bounded, $0 \le Q \le 1$, and normalized to unity with respect to the phase space measure $\prod_j \text{d}r_j \text{d}s_j/(2\pi)$.

Hence, an entropy associated to the Husimi $Q$-distribution exists, known as the Wehrl entropy~\cite{Wehrl1978,Wehrl1979}
\begin{equation}
    S_{\text{W}} (Q) = - \prod_j \int \frac{\text{d} r_j \, \text{d} s_j}{2 \pi} Q \ln Q.
    \label{eq:WehrlEntropyDefintion}
\end{equation}

The Wehrl entropy is strictly monotonous under partial trace and can therefore be regarded as a classical (differential) entropy~\cite{Haas2021}.

The Wehrl-Lieb inequality bounds the Wehrl entropy from below~\cite{Lieb1978,Lieb2014} (for $2N$ dimensional phase space)
\begin{equation}
    S_{\text{W}} (Q) \ge N,
    \label{eq:WehrlLiebEUR}
\end{equation}
and can therefore be understood as an entropic uncertainty relation in the quantum mechanical phase space. In contrast to many common uncertainty relations, it is not invariant under one-mode squeezing but is instead preserved under rotations in phase space.

It is useful to make a linear transformation to non-local EPR-type variables~\cite{Einstein1935},
\begin{align}
    r_{\pm} &= r_1 \pm r_2, &  s_{\pm} &= s_1 \pm s_2.
    \label{eq:SumDifferencesVariables}
\end{align}
By integrating over half of these variables one obtains from \eq{eq:HusimiQDefinition} marginalized distributions 
\begin{equation}
Q_{\pm} (r_{\pm}, s_{\mp})
        = \int \frac{\text{d} r \, \text{d} s}{2 \pi} Q(r,s,\mp r \pm r_{\pm}, \pm s \mp s_{\mp}).
\label{eq:marginalizeQpm}
\end{equation}
These are normalized to unity with respect to the measure $\text{d}r_\pm \text{d}s_\mp /(2\pi)$ but, because of the twisted assignment $(r_\pm,s_\mp)$, they are not themselves Husimi $Q$-distributions \footnote{The appearance of the mixed variable pairs $(r_+,s_-)$ and $(r_-,s_+)$ can be understood when employing the positive partial transpose criterion for continuous quantum variables~\cite{Simon2000}.}. One can associate to them an entropy $S_\text{M}(Q_\pm)$ that is defined in analogy to Wehrl's entropy in \eq{eq:WehrlEntropyDefintion} even though $s_\mp$ is not the conjugate momentum of $r_\pm$.

\section{Inseparability criteria}
We first derive criteria for pure states and show they generalize to mixed states. We consider pure separable states, for which the density operator is a product $\rho = \rho_1 \otimes \rho_2$. Here, the global Husimi $Q$-distribution factorizes
\begin{equation}
    Q (r_1, s_1, r_2, s_2) = Q_1 (r_1, s_1)\,Q_2(r_2, s_2),
\label{eq:HusimiProductForm}
\end{equation}
where $Q_j (r_j,s_j)$ denotes the marginals of the global Husimi $Q$-distribution. 

Inserting \eq{eq:HusimiProductForm} in \eq{eq:marginalizeQpm} yields
\begin{align}
        Q_{\pm} (r_{\pm}, s_{\mp})
        &= (Q_1 * Q_2^{(\pm)}) (r_{\pm}, s_{\mp}),
\end{align}
where $*$ denotes a convolution and $Q_2^{(\pm)} \equiv Q_2 (\pm r,\mp s)$. Invoking the two-dimensional entropy power inequality~\cite{Shannon1948b,Cover2006,Verdu2006}
\begin{equation}
    e^{S (Q_A * Q_B)} \ge e^{S(Q_A)} + e^{S (Q_B)},
\end{equation}
for any two two-dimensional Husimi $Q$-distributions $Q_A$ and $Q_B$, as well as the invariance of the Wehrl entropy under mirror reflections in phase space allows to write 
\begin{equation}
    e^{S_\text{M}(Q_{\pm})} \ge e^{S_{\text{W}} (Q_1)} + e^{S_{\text{W}} (Q_2)}.
    \label{eq:WehrlEntropyPowerInequality}
\end{equation}
Thus, we find the pair of inequalities
\begin{equation}
    S_\text{M} (Q_{\pm}) \ge \ln \left(e^{S_{\text{W}} (Q_1)} + e^{S_{\text{W}} (Q_2)} \right),
    \label{eq:CriteriaPureStatesStrong}
\end{equation}
which provide a state-dependent lower bound on the entropies $S_\text{M} (Q_{\pm})$ obeyed by all pure product states. Hence, pure states for which $S_\text{M} (Q_{\pm})$ violates this bound are necessarily entangled. We call \eq{eq:CriteriaPureStatesStrong} the \textit{strong} criteria.

To obtain a state-independent bound, we apply the Wehrl-Lieb inequality \eq{eq:WehrlLiebEUR} to both subsystems, leaving us with the criteria
\begin{equation}
    S_\text{M} (Q_{\pm}) \ge 1 + \ln 2.    
    \label{eq:CriteriaPureStates}
\end{equation}
As the latter relations are in general less tight than \eq{eq:CriteriaPureStatesStrong}, we call them the \textit{weak} criteria.

We can generalize the weak criteria to mixed states by starting with a general mixed separable state $\rho =\textstyle\sum_i p_i \left(\rho_1^i \otimes \rho_2^i \right)$, where $p_i \ge 0$ and $\sum_i p_i = 1$. On the level of the global Husimi $Q$-distributions, one has an analogous decomposition, leading, via \eq{eq:marginalizeQpm}, to
\begin{equation}
    Q_{\pm} (r_{\pm}, s_{\mp}) = \sum_i p_i \, Q^i_{\pm} (r_{\pm}, s_{\mp}).
\end{equation}
Using concavity of the entropy $S_\text{M}(Q_\pm)$~\cite{Lieb2005}, we find
\begin{equation}
    \begin{split}
        S_\text{M}(Q_{\pm}) \ge \sum_i p_i S_\text{M}(Q^i_{\pm}) \ge 1 + \ln 2,
    \end{split}
    \label{eq:CriteriaMixedStates}
\end{equation}
where we have employed the strong pure state criteria \eq{eq:CriteriaPureStatesStrong} and then the Wehrl-Lieb inequality \eq{eq:WehrlLiebEUR}. Therefore, the weak Wehrl entropic criteria for pure product states \eq{eq:CriteriaPureStates} generalize identically to mixed states. One could derive a set of strong criteria for mixed states however this requires the knowledge about the decomposition of $\rho$, which is inaccessible in experiments. The violation of inequality \eq{eq:CriteriaMixedStates} thus flags entanglement rendering it an inseparability criterion.

\section{Example states}
\subsection{Gaussian states}
An important class of states to consider is Gaussian states, which can be fully characterized by their first- and second-order moments. Since entropies are generally invariant under constant shifts of variables, we assume without loss of generality that the mean values vanish $\langle r \rangle = \langle s\rangle =0$. Hence, we only need to specify the covariance of the state
\begin{equation}
    \gamma = \begin{pmatrix}
    \braket{r^2} & \braket{rs} \\
    \braket{sr} & \braket{s^2}
    \end{pmatrix} \equiv \begin{pmatrix}
    \sigma^2_r & \sigma_{r s} \\
    \sigma_{r s } & \sigma_s^2
    \end{pmatrix},
    \label{eq:CovarianceMatrixWigner}
\end{equation}
which is also the covariance matrix of the Wigner $W$-distribution. The diagonal entries contain the variances of the corresponding marginal distributions, while the off-diagonal elements contain the covariance. One can always choose rotation angles $\vartheta_i$ such that $\sigma_{r s} = 0$, which aligns the coordinate axes along the principal axes.

For the Husimi $Q$-distribution, we define the covariance matrix as
\begin{equation}
    V_{i j} \equiv \frac{1}{2} \braket{\{u_i,u_j \}}_Q,\label{eq:husimiCovMatrix}
\end{equation}
where $u = (r,s)$ and the subscript $Q$ indicates the expectation value with respect to the Husimi $Q$-distribution. Given that the Husimi $Q$-distribution can be obtained via a Weierstrass transform of the Wigner $W$-distribution with respect to the vacuum $W_0$,
% \begin{equation}
%     Q (r,s) = 2 \pi \int \text{d} r' \, \text{d} s' \, W(r',s') W_0 (r-r',s-s'),
% \end{equation}
we can reconcile the two covariance matrices \eq{eq:husimiCovMatrix} and \eq{eq:CovarianceMatrixWigner} as $ V = \gamma + \gamma_0$. Then, expressed in terms of the twisted variables $(r_{\pm},s_{\mp})$, the Husimi $Q$-distribution of a general Gaussian quantum state leads to
\begin{equation}
    Q_{\pm} (r_{\pm},s_{\mp}) = \frac{1}{Z} e^{- \frac{1}{2} (r_{\pm}, s_{\mp})^T V_{\pm}^{-1} (r_{\pm}, s_{\mp})},
    \label{eq:GaussianHusimiQ}
\end{equation}
where $Z = \text{det}^{1/2} \, V_{\pm}$ is a normalization constant.

The entropy $S_\text{M}$ of a state with covariance matrix $V_{\pm}$ is maximized by a Gaussian distribution of the form \eq{eq:GaussianHusimiQ}, such that
\begin{equation}
    1 + \frac{1}{2} \ln \det V_{\pm} \ge S_{\text{M}} (Q_{\pm}) \ge 1 + \ln 2,
    \label{eq:CovarianceCriteria}
\end{equation}
holds for all $Q_{\pm} (r_{\pm},s_{\mp})$. Therefore, the weak entropic criteria \eq{eq:CriteriaMixedStates} imply a set of second-order based criteria and the two are equivalent for Gaussian states.

\begin{figure}[t]	
        \includegraphics[width=\columnwidth]{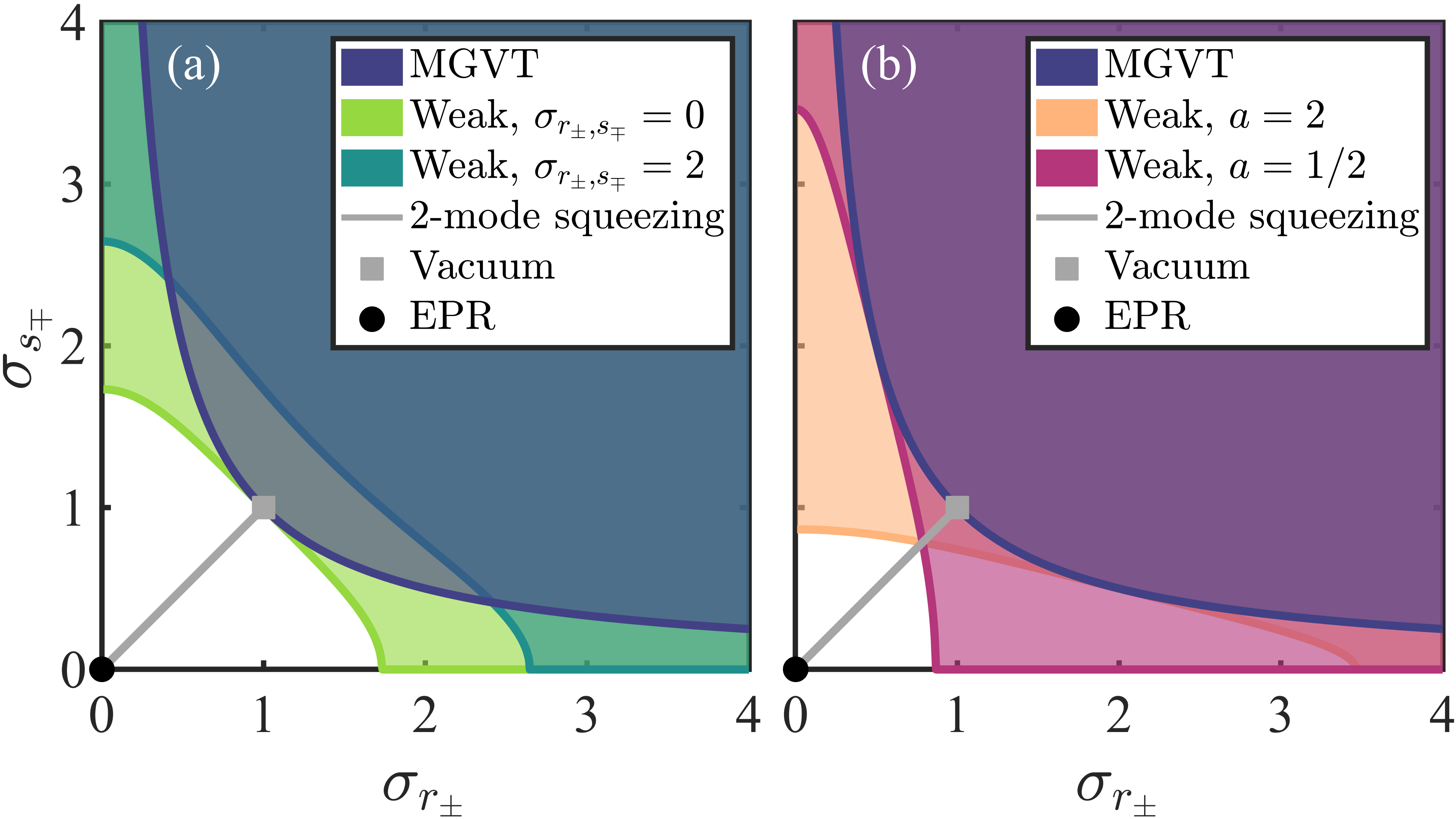}
        \caption{Regions where the MGVT criteria \eq{eq:MGVTCriteria} and the second-order criteria \eq{eq:CriteriaVariance} are fulfilled, plotted against the two marginal variances $\sigma_{r_\pm}$ and $\sigma_{s_\mp}$. The regions below the covering curves indicate entanglement. (a) We vary $\sigma_{r_\pm s_\mp}$ and fix $a =1$. The second-order criteria \eq{eq:CriteriaVariance} automatically account for an impractical alignment of the coordinate axes, while the MGVT criteria \eq{eq:MGVTCriteria} can become less tight for an improper alignment. (b) We vary $a$ and fix $\sigma_{r_\pm s_\mp} = 0$. The MGVT criteria \eq{eq:MGVTCriteria} are invariant under an equal amount of local squeezing $a$, whereas the second-order criteria \eq{eq:CriteriaVariance} must be optimized over $a$. We have additionally marked the two-mode squeezed state for all squeezing parameters $\lambda$ (gray line) from the vacuum state $\sigma_{r_{\pm}}=\sigma_{s_{\mp}}=1$ (gray square) up to the fully-correlated EPR-state $\sigma_{r_{\pm}}=\sigma_{s_{\mp}}=0$ (black dot).} \label{fig:CriteriaVariance}
\end{figure}

Our criteria are invariant under rotations (see Appendix~\ref{sec:symplectic}), however they are not invariant under local squeezing. If we consider equal amounts of local squeezing $a>0$, % such that $S = \text{diag} (a, 1/a)$, 
the second-order criteria can be rewritten as
\begin{equation}
    \left(\sigma^2_{r_\pm} + a^2 \right) \left(\sigma^2_{s_\mp} + \frac{1}{a^2} \right) \ge 4 + \sigma^2_{r_\pm s_\mp}.
    \label{eq:CriteriaVariance}
\end{equation}
After optimizing over the local squeezing parameter $a$ and choosing angles $\vartheta_i$ such that the coordinate axes are parallel to the principal axes (i.e., $\sigma^2_{r_\pm s_\mp} =0$) these criteria are equivalent to the MGVT criteria~\cite{Mancini2002}
\begin{equation}
    \sigma_{r_\pm} \sigma_{s_\mp} \ge 1,
    \label{eq:MGVTCriteria}
\end{equation}
which themselves are equivalent to the entropic criteria in Ref.~\cite{Walborn2009} for Gaussian states. In contrast to our second-order criteria \eq{eq:CriteriaVariance}, the MGVT criteria are invariant under equal amounts of local squeezing. However, they are not invariant under rotations in the $\pm$-phase space as they do not contain the covariances. In this sense, the two second-order criteria \eq{eq:CriteriaVariance} and \eq{eq:MGVTCriteria} behave complementary under rotations and squeezing. This is summarized in \fig{fig:CriteriaVariance}. 

Note that both second-order criteria \eq{eq:CriteriaVariance} and \eq{eq:MGVTCriteria} are only sufficient criteria for inseparability. This implies that the criteria by Simon~\cite{Simon2000} and Duan \textit{et al.}~\cite{Duan2000} (after optimization over local squeezing parameters and angles) are generally stronger in the Gaussian regime.

\subsection{Non-Gaussian states}
To exemplify the strengths of our entropic criteria, we consider a set of non-Gaussian entangled states that cannot be witnessed by second-order criteria to test the weak \eq{eq:CriteriaMixedStates} and strong criteria \eq{eq:CriteriaPureStatesStrong}~\footnote{When the state is pure, however, the Wehrl mutual information already provides a perfect entropic witness~\cite{Haas2021}}.

First, we consider the planar $N00N$ states that are given by
\begin{equation}
    \ket{\psi_N} = \frac{1}{\sqrt{2 (1 + \delta_{0 N})}} \left(\ket{N,0} + \ket{0,N} \right),
    \label{eq:N00NStatesMain}
\end{equation}
with $N \in \mathbb{N}_0$.

\begin{figure}[t]	
         \includegraphics[width=\columnwidth]{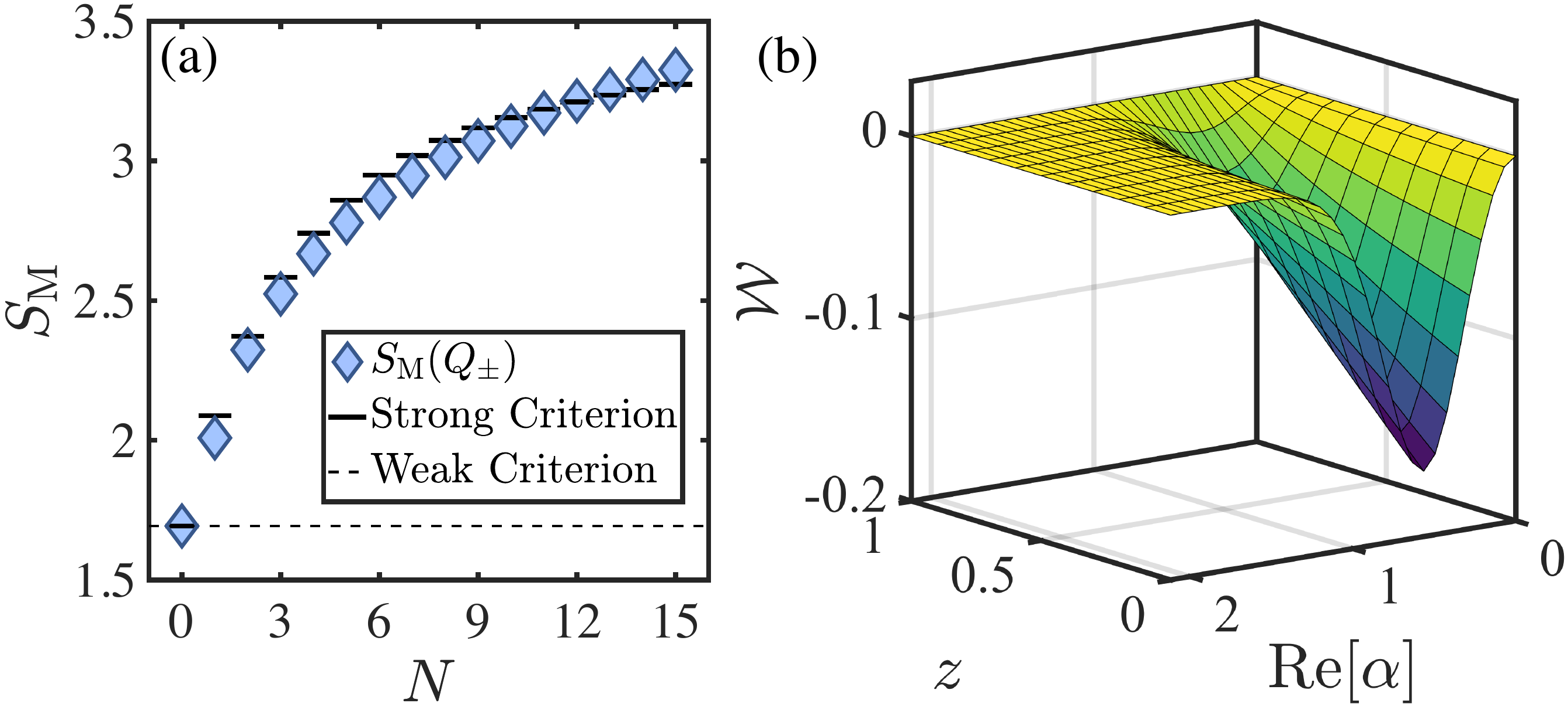}
        \caption{(a) Wehrl entropic criteria for the first 15 $N00N$ states. Entanglement is witnessed when the points (blue diamonds) are below the bars. The strong criterion \eq{eq:CriteriaPureStatesStrong} witnesses entanglement up to $N=11$, while the weak criterion \eq{eq:CriteriaMixedStates} does not witness any entanglement. (b) Testing the weak criterion \eq{eq:CriteriaMixedStates} for the dephased cat state \eq{eq:CatState} with $s=0$. Negative $\mathcal{W} = S_\text{M}(Q_\pm)-1-\ln2$ witnesses entanglement. Entanglement is detected for $\mathrm{Re}[\alpha]>0$ and $z<1$, but the difference $\mathcal{W}$ converges quickly to zero for $\mathrm{Re}[\alpha] \gtrsim 3/2$. In both cases $\vartheta_1 = \vartheta_2 = 0$.} \label{fig:CatState}
\end{figure}

We plot the behavior for the two criteria in \fig{fig:CatState}(a) up to $N=15$ for $Q_{+} (r_+, s_-)$. Our strong Wehrl criteria \eq{eq:CriteriaPureStatesStrong} witnesses entanglement up to $N=11$. This goes beyond the capabilities of entropic criteria based on marginal distributions. For example, the witness in Ref.~\cite{Walborn2009} detects entanglement up to $N=5$, while the generalization in Ref.~\cite{Saboia2011} is capable of certifying entanglement up to $N=6$. The weak criteria \eq{eq:CriteriaPureStates} do not witness any entanglement, which is analogous to the results in Refs.~\cite{Walborn2009,Saboia2011}.

As a second example, we consider the Schr\"{o}dinger cat state
\begin{equation}
    \begin{split}
        \rho &= N(\alpha) \Big[\ket{\alpha,\alpha} \bra{\alpha, \alpha} + \ket{- \alpha, - \alpha} \bra{-\alpha, -\alpha} \\
        &- (1-z) \left( \ket{\alpha, \alpha} \bra{-\alpha,-\alpha} + \ket{-\alpha, -\alpha} \bra{\alpha,\alpha} \right) \Big],
    \end{split}
    \label{eq:CatState}
\end{equation}
where $0 \le z \le 1$ and $N(\alpha) =  (1 + (1-z) e^{- 4 \abs{\alpha}^2})/2$ normalizes the state. For $z=0$, \eq{eq:CatState} is a pure Schr\"{o}dinger cat state and for $z>0$ it is a dephased cat state that is mixed.

In \fig{fig:CatState}(b), we show that entanglement is witnessed for all values of $\mathrm{Re}[\alpha] > 0, \mathrm{Im}[\alpha]=0$ and $z<1$ by the weak criteria \eq{eq:CriteriaMixedStates}. In principle, detecting entanglement in \eq{eq:CatState} depends on $\mathrm{Im}[\alpha]$ too. However one can choose arbitrary $\vartheta_i$ such that the optimal $\alpha$ only depends upon its real component. The inseparability criteria in \eq{eq:CriteriaMixedStates} are violated most in the region $0\lesssim \mathrm{Re}[\alpha]\lesssim3/2$, while for larger $\mathrm{Re}[\alpha] \gtrsim 2$, the difference between a superposition and a mixture becomes suppressed exponentially. In contrast, the entropic criteria in Refs.~\cite{Walborn2009,Saboia2011} certified entanglement only for $\text{Re}[\alpha] \gtrsim 5/3$ and $z<1$ when using $\vartheta_1 = \vartheta_2 = 0$.

% \textit{Outlook}---\red{here we can discuss a bit about experimental applications.}
\section{Possible experimental realizations}
%general protocol
The protocol for applying our entropic witnesses in experiments is to measure the full Husimi $Q$-distribution \eq{eq:HusimiQDefinition} and to estimate from the obtained data the entropies of the EPR-type variables \eq{eq:SumDifferencesVariables}.
%possible ways to measure Husimi (tomographic and heterodyning)
Techniques for measuring Husimi $Q$-distributions are well established in quantum optics and include (i) tomographic schemes applying displacements to the prepared states before measuring its vacuum projection~\cite{Lvovsky2009} and (ii) heterodyne measurements~\cite{Noh1991}.
%Experiments that have realized it
Recently, these schemes have been realized in other experimental systems including ultracold spinor Bose gases~\cite{Kunkel2021}, cold atoms in cavities~\cite{haas2014,barontini2015}, and circuit QED architectures~\cite{kirchmair2013}. Additionally, measurements of the $Q$-distribution via coherent displacements and measurements of the vacuum state~\cite{Landon2018} could readily be realized in trapped-ion systems~\cite{gaerttner2017}. The works listed here have measured (or have the potential to measure) Husimi distributions for a monopartite system (i.e., a single mode). Therefore, further work would need to be carried out to extend measurements to a bipartite system (i.e., two modes) so that the witness presented here could be applied.
%Advantages wrt marginal detection

Both schemes could carry practical advantages--particularly for cold atom systems--compared to the detection of marginals: Scheme (i) overcomes the problem that high detector resolution, necessary for accurate entropy estimation, by requiring only the technically easier and more scalable task of detecting the probability of all particles being in the same state or mode (vacuum detection). Both schemes avoid determining the detection angles $\vartheta_i$ in \eq{eq:RotatedVariables}, which is costly in terms of experimental runs in cold atom experiments. Additionally, tomography angles are often difficult to control precisely here. Due to these features, our entanglement criteria will potentially enable the experimental certification of entangled states beyond the reach of currently available methods.

%Disclaimer/outlook: 
We note that our derivation uses the Husimi $Q$-distribution with respect to the harmonic oscillator coherent states. This accurately describes the distribution obtained from heterodyne measurements in quantum optics, however, for many of the aforementioned experimental platforms, the applicability of this description is limited due to finite particle numbers and will require a generalization to SU(2) coherent states. Additionally, the extraction of entropies from experimental data is a challenging task in the presence of finite detector resolution and statistical noise~\cite{Schneeloch2019}. The required measurement statistics for a given experimental platform, prepared state and suitable entropy extraction scheme need to be evaluated carefully in order to make statements about the actual experimental cost and feasibility. These aspects are subject to ongoing and future research in our group.

\section{Conclusions}
We have derived inseparability criteria in terms of variants of the Wehrl entropy, which can be applied when measuring the Husimi $Q$-distribution. In contrast to most (entropic) criteria, we have shown that our criteria are invariant under rotations in phase space while depending on the local squeezing parameters. As a consequence, the criteria witnessed some entangled states that are undetectable using entropic criteria based on marginal distributions. We have discussed the implementation of our witness for a wide variety of experimental platforms and expect it to perform strongly in comparison with previous marginal criteria. Future theoretical studies should generalize the presented approach to spin operators fulfilling a SU(2) algebra to formulate entropic criteria for discrete quantum spin systems. 

%incorporate memory, possibly leading to steering inequalities in terms of a conditional Wehrl entropy. %Furthermore, it would be of great interest to derive inseparability for SU$(2)$ coherent states. This would broaden the range of applicability of our inseparability criteria to a larger range of experimental platforms and classes of preparable states.

\begin{acknowledgements}
We thank Markus Oberthaler and Markus Schr\"{o}fl for useful discussions. This work is supported by the Deutsche Forschungsgemeinschaft (DFG, German Research Foundation) under Germany's Excellence Strategy EXC 2181/1 - 390900948 (the Heidelberg STRUCTURES Excellence Cluster) and under SFB 1225 ISOQUANT - 273811115 as well as FL 736/3-1.
\end{acknowledgements}

\begin{appendix}

\section{Symplectic transformations}\label{sec:symplectic}

To validate our second-order criteria, we consider how symplectic transformations in the $\pm$-variables affect the Husimi $Q$-distribution. A general symplectic transformation $S \in \text{Sp}(2,\mathbb{R})$ fulfilling $S^T \Omega S = \Omega$, with $\Omega$ being the symplectic form, can easily be applied to the original Wigner $W$-distribution~\cite{Simon2000}. This causes the corresponding covariance matrix $\gamma$ to transform as 
\begin{equation}
    \gamma \to \gamma' = S \gamma S^T.
\end{equation}
In contrast, the distribution $Q_{\pm}$ does not transform in a straight forward manner. We therefore restrict our analysis of the symplectic group to only Gaussian states. The second-order criteria \eq{eq:CovarianceCriteria} then transform as
\begin{equation}
    \begin{split}
        \det V_{\pm} \to \det V'_{\pm} &= \det \left(S \gamma_{\pm} S^T + \gamma_0 \right) \\ 
        &= \det \left(\gamma_{\pm} + \gamma_0 (S^T S)^{-1} \right),
    \end{split}
\end{equation}
where we used $\det S = \det S^T = 1$ and that the vacuum covariance matrix is the identity $\gamma_0 = \mathds{1}$. This shows that invariance of $\det V_{\pm}$ is equivalent to $S$ being an orthogonal matrix $S^T S = \mathds{1}$ corresponding to a rotation. Therefore, the orientation of the axes is unimportant for the analysis of entanglement. This result generalizes to arbitrary marginals of Husimi $Q$-distributions since any two-dimensional (differential) entropy is invariant under a rotation.

\section{Explicit Husimi $Q$-distributions}

Our first example in the main text was the set of $N00N$ states, given in \eq{eq:N00NStatesMain}. The global Husimi $Q$-distribution for $\vartheta_1 = \vartheta_2 = 0$ is
\begin{equation}
    \begin{split}
        Q (r_1,s_1,r_2,s_2) &= \frac{e^{-\frac{1}{2} \left(r_1^2 + s_1^2 + r_2^2 + s_2^2 \right)}}{2^{N+1} \, N! \, (1 + \delta_{0 N})} \\
        & \times \left( (r_1 - i s_1)^N + (r_2 - i s_2)^N \right) \\ 
        & \times \left( (r_1 + i s_1)^N + (r_2 + i s_2)^N \right).
    \end{split}
    \label{eq:N00NStatesHusimiQ}
\end{equation}

The second example that we considered was the Schr\"{o}dinger cat state given in \eq{eq:CatState}. Recall that for $z=0$, \eq{eq:CatState} is a pure Schr\"{o}dinger cat state and for $z>0$ it is a dephased cat state that is mixed. The full Husimi $Q$-distribution for $\vartheta_1 = \vartheta_2 = 0$ is
\begin{equation}
    \begin{split}
        &Q (r_1, s_1, r_2, s_2) \\
        &= N(\alpha) \Big[e^{- \frac{1}{2} \left((r-r_1)^2 + (s-s_1)^2 + (r-r_2)^2 + (s-s_2)^2 \right)} \\
        &\hspace{1.6cm}+ e^{- \frac{1}{2} \left((r+r_1)^2 + (s+s_1)^2 + (r+r_2)^2 + (s+s_2)^2 \right)} \\
        &\hspace{1.6cm}+ 2(1 - z) \, e^{-r^2 - s^2 -\frac{1}{2} \left(r_1^2 + s_1^2 + r_2^2 + s_2^2 \right)} \\
        &\hspace{1.6cm}\times \cos \left( r(s_1 + s_2) - s (r_1 + r_2) \right) \Big],
    \end{split}
    \label{eq:HusimiQCatState}
\end{equation}
where we use the parameterization $\alpha = (r+is)/\sqrt{2}$.
\end{appendix}

%%%%%%%% Bibliography %%%%%%%%

\bibliography{references.bib}
\end{document}